# A Platform for Addressing Individual Magnetite Islands Grown Epitaxially on Ru(0001) and Manipulating Their Magnetic Domains

Sandra Ruiz-Gómez, Eva María Trapero, Claudia Fernández-González, Adolfo del Campo, Cecilia Granados-Miralles, José Emilio Prieto, Muhammad Waqas Khaliq, Miguel Angel Niño, Michael Foerster, Lucía Aballe, and Juan de la Figuera*

**ABSTRACT:** We have grown high-quality magnetite micrometric islands on ruthenium stripes on sapphire through a combination of magnetron sputtering (Ru film), high-temperature molecular beam epitaxy (oxide islands), and optical lithography. The samples have been characterized by atomic force microscopy, Raman spectroscopy, X-ray absorption and magnetic circular dichroism in a photoemission microscope. The magnetic domains on the magnetite islands can be modified by the application of current pulses through the Ru stripes in combination with magnetic fields. The modification of the magnetic domains is explained by the Oersted field generated by the electrical current flowing through the stripes underneath the magnetite nanostructures. The fabrication method is applicable to a wide variety of rock salt and spinel oxides.

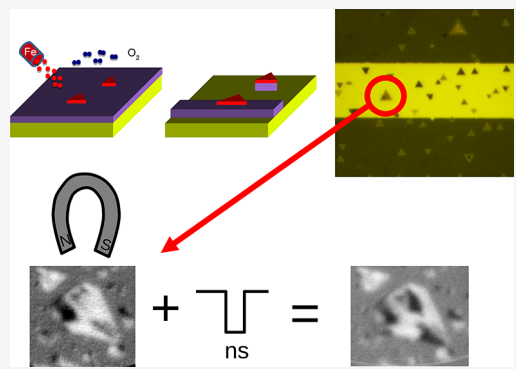

## INTRODUCTION

The quality of materials can severely impact their properties. This truism has been thoroughly proven by the microelectronics industry, where the current capabilities lean on the ability to grow compositionally controlled materials with extremely low defect densities. In other areas, there is still a great deal of margin for improvement. Such is the case of spintronics, where often polycrystalline materials are used. This is reasonable, as many properties are averaged over larger scales. For example, the magnetic domain walls are often thicker than the polycrystalline grain size. However, there are examples in which this is not the case. For example, skyrmion motion is affected by defects[1] and the ability to confine and control spin waves at the edges of nanostructures is likely to require atomically perfect materials.[2]

In the past few years we have explored the growth of several ferrimagnetic and antiferromagnetic oxides of high crystalline quality by means of oxygen-assisted high-temperature molecular beam epitaxy on single-crystal Ru(0001) substrates.[3] This method has been successfully used to obtain atomically flat micrometer-wide and nanometer-high triangular islands of several ferrimagnetic spinel ferrites, among them iron spinel (i.e., magnetite),[4] cobalt ferrite,[5] and nickel ferrite,[6] as well as antiferromagnetic $Ni_xCo_{1−x}O$[7] and $Ni_xFe_{1−x}O$ islands of similar quality,[8] in addition to rare-earth oxides such as ceria[9] and praseodymium.[10] The magnetic oxides grown in such a way present magnetic domains in remanence which are orders of magnitude larger than those typically observed in thin films. This is attributed in part to the lack of antiphase boundaries,[11] as each of the islands arises form a single nucleus and the growth process is stopped before coalescence. In the case of magnetite, magnetic closure domains dictated by shape anisotropy have been observed[4] and modified through the application of external magnetic fields.[12,38]

The use of single-crystal bulk Ru(0001) substrates can be substituted by thin Ru films deposited on insulating substrates, as proved by the growth of ceria[13] and graphene[14] on such films. We have recently characterized those thin films as substrates[15] and found them to be of a quality comparable to that of bulk single crystals. Furthermore, the quality of oxide islands grown on the films is similar to that of those grown on single crystals. In particular, magnetite crystals show a Verwey transition as detected by Raman spectroscopy.[16]

One advantage of such Ru films is that they can be removed by standard etchants developed by the microelectronics industry.[17] We thus propose that a powerful platform for the implementation of electrical control of high-quality nanostructures of magnetic oxides is their growth by high-temperature oxygen-assisted molecular beam epitaxy on thin films of ruthenium, with a final step of optical lithography to define conductive paths on an otherwise insulating substrate. In the present work, we present validation for such a platform for the specific case of magnetite islands. After growing them on thin Ru films as substrates, we lithographically defined stripes in the metal. We then checked that the magnetite islands were unaffected by the procedures of etching and removing the resist. Finally we demonstrated the modification of the magnetic domains of the nanostructures by several means, as observed by X-ray magnetic circular dichroism (XMCD) in photoemission microscopy (PEEM), thus validating our approach.

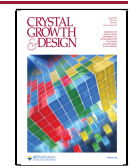

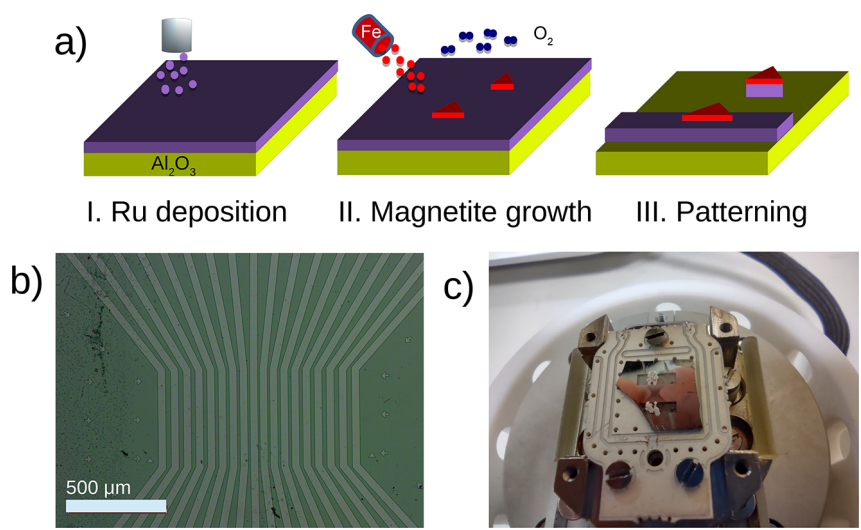

**Figure 1.** (a) Schematic of the growth procedure. From left to right: (I) growth of a Ru film by magnetron sputtering, (II) growth of magnetite islands by high-temperature oxygen-assisted molecular beam epitaxy, and (III) lithographic definition of stripes. (b) Optical microscopy of the sample with the developed resist on top. (c) Sample mounted in the holder that allows the application of external magnetic fields.

## EXPERIMENTAL METHODS

Ru films have been grown, following reports by several groups that indicated epitaxial growth,[13,14,18] by direct-current magnetron sputtering in a home-made sputtering chamber with a base pressure of $10^{-6}$ mbar. The sapphire substrates, with the (0001) orientation, 99.998% pure, and polished to 0.3 nm, were provided by Siegert Wafer. The growth, using 99.95% Ru targets from Evochem, was performed with a typical power of 30 W over 5–10 min with a sample to target distance of 10 cm. The sapphire substrates were heated to 500 °C before and during the growth.

Magnetite islands have been grown in ultrahigh-vacuum chambers equipped with low-energy electron microscopy (LEEM), which permits the observation of the growth front in real time. We have used the Elmitec LEEM III instrument at the Instituto de Química Física Blas Cabrera as well as a similar instrument at the CIRCE station of the Alba synchrotron.[19] The samples were degassed at temperatures up to 1000 °C. After such a procedure the Ru films usually present a sharp (1 × 1) low-energy electron diffraction (LEED) pattern that corresponds to a well-ordered Ru(0001) surface.[13] In cases where other LEED patterns were observed, a mild sputtering cycle with Ar ions at 1 keV was performed, and the annealing step was repeated. The magnetite islands were grown following our established protocol,[4,20,21] by depositing iron in a background molecular oxygen pressure of $1 \times 10^{-6}$ mbar at a substrate temperature of 900 °C. The iron source was a 5 mm diameter iron rod heated by electron bombardment within a water jacket. Typical deposition rates were about 15 min per Fe layer.

The samples were spin-coated with a high-contrast AZ 1512 HS photoresist with a typical thickness of 1.2 μm. A lithographic pattern was defined by exposing to light in a laser lithography system and developed with a multistriped pattern of narrow channels 10–30 μm wide and 500 μm long which gradually widened over 1 mm to a width of 100 μm.

To etch the noncovered areas, we employed a solution of 9 wt % of acetic acid and 22 wt % ceric ammonium nitrate in water.[14,17]

After the growth of the magnetite islands, the sample was coated with a photoresist, and a stripe pattern was projected onto the sample and developed (the protocol is summarized in Figure 1a). The etchant used, designed for Ru, efficiently removed the Ru in the exposed regions. However, the etchant also damaged the resist in the covered areas to the point that solvents such as acetone and pyrrolidone were not effective in removing it. Instead, we have successfully used piranha solution ($H_2SO_4 + H_2O_2$) to such an end.

The samples, one of which is shown in Figure 1, were contacted by wire-bonding with 100 μm Al wire to a printed circuit board (PCB) mounted in an Alba sample holder which included a coil for generating a magnetic field in the plane of the PCB.[22] The resistance of the individual stripes was about 300 Ω, which is in good agreement with the Ru resistivity[23] and the channel geometry considering a Ru stripe height of 10 nm.

A setup for generating and measuring current pulses was mounted inside the high-voltage rack of the PEEM microscope. It included a 40 V/60 ns pulse generator from AVTECH electrosystems, Model AVI-MP-P, a custom-made polarity switch box, the device itself, and a 50 Ω resistor to ground. The shape of the pulses both before the device and between the 50 Ω resistor and the device was monitored with a Tektronics TPS 2048S digital oscilloscope. The pulses were applied with the high voltage of the microscope switched off to prevent damage to the stripes.

## RESULTS AND DISCUSSION

The Ru films grown by magnetron sputtering on $Al_2O_3$(0001) single crystals usually have a high density of steps, although some authors like Sauerbrey et al.[13] reported that such films can have even a lower density of steps than well-prepared single crystals.

Upon Fe deposition under an oxygen atmosphere, the Ru film is first covered by a FeO(111) wetting layer whose thickness is two atomic layers for the conditions we used,[24–28] followed by the growth of 3-dimensional magnetite islands with thicknesses ranging from a few nanometers to a hundred nanometers and a lateral extension of several micrometers.[20]

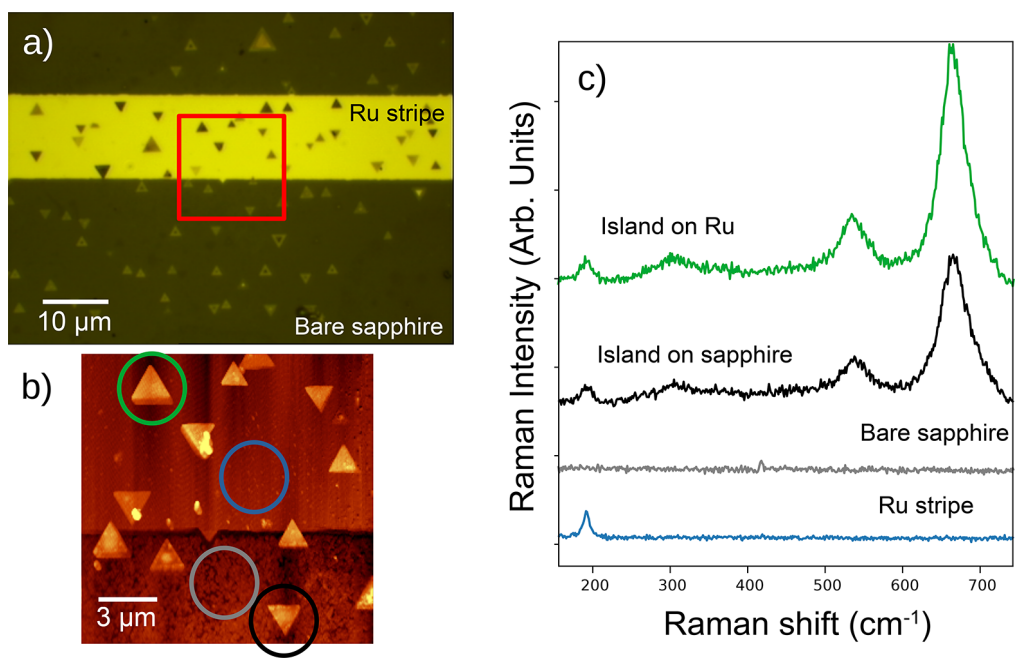

**Figure 2.** (a) Optical microscopy image of a Ru stripe with magnetite triangular islands. (b) Atomic force microscopy images of the area marked in (a) with a red square. The locations where the Raman spectra in c) were acquired are marked with circles of the same color as each spectrum. (c) Raman spectra acquired in the different regions of the film. From bottom to top: spectra on a ruthenium stripe (blue), on the bare sapphire (gray), on an island on sapphire (black), and on an island on the Ru stripe (green).

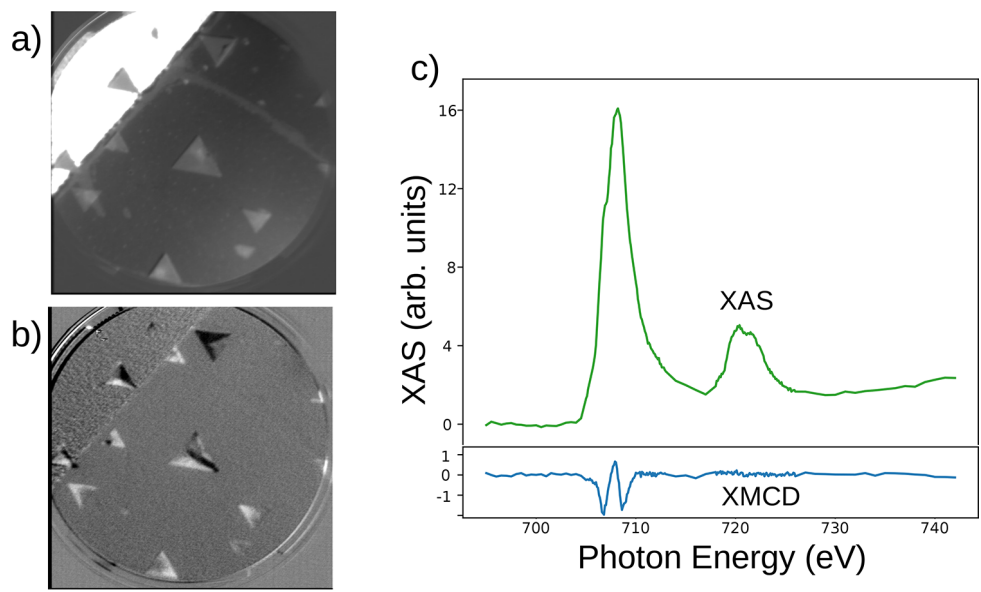

**Figure 3.** (a) XAS spectroscopy image of a Ru stripe and the sapphire around it, acquired at a photoelectron energy near the maximum of the $L_3$ white line. The field of view is 10 μm. (b) The same area presenting the XMCD image. (c) XAS and XMCD spectra acquired on a single domain of a magnetite island.

Depending on the particular details of the growth temperature and postprocessing of the samples (like further annealing steps), the magnetite islands either extend down to the Ru substrate or sit on top of the FeO wetting layer. Our goal here is to establish that the islands correspond to magnetite, so we refer the reader to the published works on this subject.[21,26] The density of islands on the Ru films is comparable to that of magnetite islands grown in single-crystal Ru substrates. It is likely that the presence of the wetting layer decouples the nucleation of the magnetite islands from the local density of the steps to some extent. We have also tried to grow the oxide islands on prepatterned substrates. However, we obtained a much higher density of smaller islands, likely due to contamination introduced during the processing steps.

The resultant devices were characterized by atomic force microscopy, by Raman spectroscopy and by X-ray absorption spectroscopy (XAS) in PEEM. The microscope images presented in Figure 2a,b show triangular islands on stripes which are elevated by 13 nm, while the islands on top have a typical thickness of 30 nm. Presumably, the elevated areas correspond to the Ru stripes on which the triangular islands (magnetite crystals) have grown and the deeper regions correspond to the bare sapphire areas. The islands present a range of different heights, from 20 nm to more than 40 nm. However, there are also islands detected directly on the sapphire.

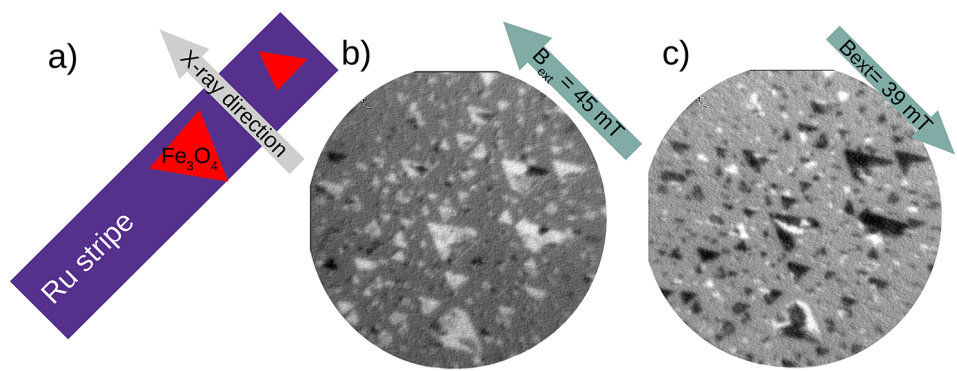

**Figure 4.** (a) Schematic of the directions of the Ru stripes relative to the applied external magnetic field, and the direction of the X-ray beam. (b, c) XMCD-PEEM images after applying a magnetic field of 45 mT (b) or 39 mT (c) in the directions indicated in each image. The field of view of the images is 10 μm.

Raman spectra are shown in Figure 2c. They were acquired respectively on the Ru stripe outside of any island (blue spectrum), on the bare sapphire areas outside of the islands (gray spectrum), on an island on the bare sapphire (black spectrum), and on an island on the Ru stripe (green spectrum). The sapphire areas only show a very small peak near 420 cm$^{-1}$, characteristic of single-crystal $\alpha$-Al$_2$O$_3$.[29] The Ru stripe shows a peak at 192 cm$^{-1}$ which corresponds to the $E_{2g}$ transverse optical phonon from the shear motion of the two sublattices of the hcp structure.[30] All of the islands present several peaks, the most prominent of which is that at 665 cm$^{-1}$. Two other peaks appear at 310 and 535 cm$^{-1}$. All these peaks arise from the magnetite spinel structure and are assigned to the $A_{1g}$ mode, and to two of the $T_{2g}$ modes.[16,31,32] This is an indication that the lithographic steps did not destroy the magnetite structure of the islands. This is true not only for the islands that were protected during the etching of the Ru (green spectrum) but also for the islands that were exposed in order to remove the Ru film (black spectrum).

The fact that the magnetite spinel modes are detected on both type of islands proves that the magnetite islands not only survived the brief piranha immersion employed to remove the remains of the resist but also, in the case of islands sitting on sapphire, survived the Ru etching agent. On the other hand, that the Ru mode is detected in both types implies that whether covered by the resist or by the magnetite islands, the underlying Ru is protected. In the former case, this is the intended outcome and it is achieved by using a resist with a thickness of over 1 μm. However, in the latter case, this indicates that magnetite is also effective in protecting the underlying Ru even if the islands are a few tens of nanometers thick. This is consistent with the islands on the bare sapphire being apparently thicker than those on Ru by an amount similar to the Ru thickness.

Whether the Ru underlying the magnetite islands survives might be related to the particular etching times employed. The islands sitting on sapphire in another sample, which was etched for a longer time, did not show the Ru peak underneath.

The devices have also been characterized by XAS, using the iron $L_{3,2}$ absorption edges, as shown in Figure 3. In the XAS image, triangular-shaped magnetite islands are detected both on the Ru stripes and on the sapphire substrate. The Ru stripe appears dark in Figure 3a due to work function differences. The islands present sides oriented along the compact directions of the Ru(0001) surface, as ascertained by LEED (not shown). In the images, the Ru stripes are aligned at 45° with respect to the horizontal direction. In Figure 3c we show the averaged XAS spectrum acquired at a magnetic single domain of a triangular island with opposite circular polarizations of the X-rays, together with the corresponding XMCD spectrum.

We first note that the XAS spectrum is characteristic of magnetite,[33] thus confirming the observation by Raman spectroscopy that the magnetite islands have survived all the steps involved in the lithography process. We also note that the XAS observation also implies that there is not a large "dead" surface layer in the magnetite islands. While the Raman signal contains the contribution of the full thickness of the islands, which is in the range of tens of nanometers, XAS-PEEM is far more surface sensitive. While the precise mean free path of very low-energy electrons in oxides applicable to our measurement kinetic energy of 1 eV is still under discussion, experimental values in magnetite are in the range from 5 nm[34] to 1.4 nm.[35]

To image the magnetic domains we make use of the XMCD effect: XAS images were acquired with opposite circular polarizations and then subtracted pixel-by-pixel. There are several energy ranges at the $L_3$ edge which provide magnetic contrast, as shown in Figure 3c. The XMCD spectrum consists of two negative peaks separated by a positive one. The former are considered to arise mainly from Fe$^{2+}$ and Fe$^{3+}$ in octahedral positions, respectively, while the latter corresponds to Fe$^{3+}$ in a tetrahedral environment. The XMCD images presented in this work have been recorded at the first negative peak. Thus, they map the local magnetization in magnetite, which corresponds to the direction of the magnetic moment of the Fe$^{2+}$ iron atoms at octahedral positions. The magnetic contrast image of the region displayed in Figure 3a is shown in Figure 3b. The images acquired provide only the component of the magnetization along the X-ray beam incidence direction, which is orthogonal to the stripe orientation (measuring along different azimuthal sample orientations can be performed to allow the reconstruction of the full magnetization vector[4]). Thus, the white areas correspond to domains with the magnetization pointing along the incoming X-ray beam direction and black ones in the opposite direction. Gray areas indicate either no net magnetization or a magnetization along a direction

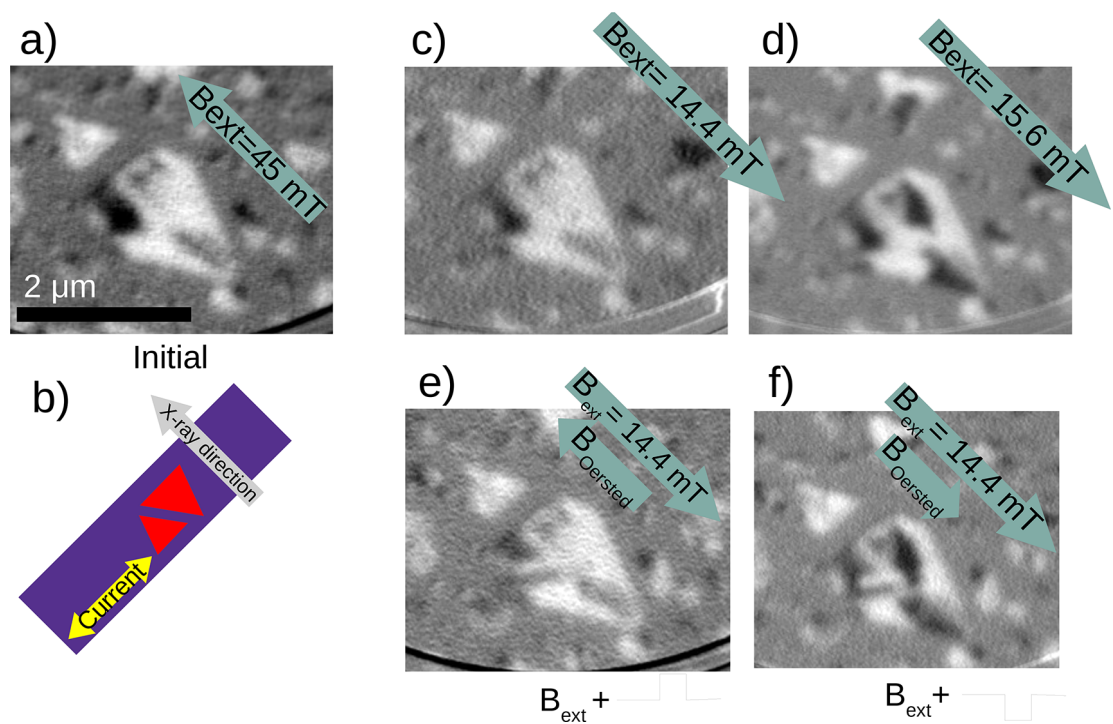

**Figure 5.** (a) Initial configuration obtained after applying a magnetic field of 45 mT in the direction indicated. (b) Schematic of the stripe. (c) Image after applying in (a) a magnetic field of 14.4 mT in the opposite direction (no changes are observed in the large island). (d) After applying a slightly higher field (15.6 mT) changes are detected in the large island. (e) Image acquired after simultaneously applying a magnetic field of 14.4 mT and one 60 ns pulse of positive polarity. No changes are observed. (f) Configuration after applying the same magnetic field and a negative polarity pulse. Changes are observed, similar to those with a higher magnetic field (d).

perpendicular to the incoming X-ray direction: i.e., along the stripe axis.

To check the possibility of manipulating the magnetic domains with an external magnetic field, we mounted a device in a special sample holder with a mini-electromagnet that allows application of an in-plane magnetic field.[22] The geometry and directions of the X-ray beam relative to the applied magnetic field are shown in the schematic in Figure 4. We note that, as we have discussed for spinel islands grown on Ru(0001), there are two different types of domain patterns: for very thin islands, the domain walls are often pinned by the defects induced by the substrate steps,[36] while taller islands tend to have domains governed by shape anisotropy.[4] The saturation magnetization of the magnetite islands grown on Ru is expected to be of a magnitude similar to that of the bulk material, as estimated from comparisons of the domain wall width with micromagnetic simulations.[4] Regarding the magnetic fields required to modify their domains, to the best of our knowledge there are no measurements of hysteresis cycles of individual islands. Nanometer thick magnetite films on Ru(0001) have a reported coercive field around 30 mT.[37] Our own research on magnetite islands grown on bulk Ru crystals has shown that fields of a few mT are enough to modify the shape anisotropy patterns in a reversible way[38] and that fields of 40−50 mT are needed to modify their magnetization patterns in remanence.[12] The image in Figure 4b corresponds to the domains observed after applying a magnetic field of 45 mT,[12,22] while the next panel shows the domains observed after reversing the applied magnetic field. Many of the smaller islands observed are single-domain, while the larger ones contain several magnetic domains. Even in the latter case, the majority of the domains of each island point in the direction of the applied field. After confirming the effect of applying an external magnetic field to the magnetite islands, we first "reset" the magnetic configuration by applying a large magnetic field as shown in Figure 4b, obtaining the initial configuration shown in Figure 5a. We then find the highest magnetic field that does not modify the magnetic configuration of the larger island (Figure 5c) and the lowest field that does modify the configuration (Figure 5d). We found that in the central island a field larger than 15 mT is required to change the domains. Finally, we use the former as a bias and apply electrical pulses flowing underneath the island. For a current of 0.13 A, the estimated Oersted field is 4 mT. Depending on the polarity of the pulses, the orientation of the Oersted field changes. If the Oersted field and the magnetic field applied by the sample holder coil are antiparallel so that the net field is smaller, no change is observed in the island (Figure 5e). If they are parallel, they add, the combined magnetic field is above the 15 mT threshold, and changes in the domains are observed (Figure 5f). The current density produced by the pulse is estimated from the total current and the stripe dimensions to be $6 \times 10^{11}$ A/m$^2$. The Oersted field is the most straightforward method of modifying the magnetic domains in a nanostructure on top of a conductive stripe.[39,40] In this sense, our results are not unexpected. However, we stress that the main point we present is that magnetic domains can be switched in a device where the shape of the nanostructures is defined by growth and not by methods such as focused ion beam or lithography and thus offers the promise of obtaining nanostructures with atomically perfect edges for future studies of magnetic domain manipulation in epitaxial oxide materials.

## ■ CONCLUSIONS

We have grown high-quality magnetite islands on a Ru film deposited on sapphire and we have defined stripes lithographically on the system. The magnetite islands survive the etching process, as proved by microspot Raman spectroscopy and atomic force microscopy. The islands also show the expected X-ray absorption spectra of magnetite. Magnetic domains can be observed on them by XMCD-PEEM. The domains can be modified by the application of an external magnetic field of a magnitude similar to that required for islands grown on Ru single crystals. Injecting current through the Ru stripes underneath the magnetite islands produces changes in selected islands when the Oersted field of the current pulses adds to the applied external magnetic field. This validates the oxide islands grown by molecular beam epitaxy on Ru films deposited on sapphire as an excellent platform to study the switching effects of currents on oxide structures, both ferrimagnetic and (in the future) antiferromagnetic.

## ■ AUTHOR INFORMATION

**Corresponding Author**

Juan de la Figuera − *Instituto de Química Física Blas Cabrera (IQF), CSIC, Madrid 28006, Spain;* orcid.org/0000-0002-7014-4777; Email: juan.delafiguera@csic.es

**Authors**

Sandra Ruiz-Gómez − *Max-Planck-Institut für Chemische Physik fester Stoffe, Dresden 01187, Germany*

Eva María Trapero − *Instituto de Química Física Blas Cabrera (IQF), CSIC, Madrid 28006, Spain*

Claudia Fernández-González − *Max-Planck-Institut für Chemische Physik fester Stoffe, Dresden 01187, Germany*

Adolfo del Campo − *Instituto de Cerámica y Vidrio, CSIC, Madrid 28049, Spain*

Cecilia Granados-Miralles − *Instituto de Cerámica y Vidrio, CSIC, Madrid 28049, Spain;* orcid.org/0000-0002-3679-387X

José Emilio Prieto − *Instituto de Química Física Blas Cabrera (IQF), CSIC, Madrid 28006, Spain;* orcid.org/0000-0003-2092-6364

Muhammad Waqas Khaliq − *Alba Synchrotron Light Facility, Barcelona 08290, Spain*

Miguel Angel Niño − *Alba Synchrotron Light Facility, Barcelona 08290, Spain;* orcid.org/0000-0003-3692-147X

Michael Foerster − *Alba Synchrotron Light Facility, Barcelona 08290, Spain*

Lucía Aballe − *Alba Synchrotron Light Facility, Barcelona 08290, Spain;* orcid.org/0000-0003-1810-8768

Complete contact information is available at:
https://pubs.acs.org/10.1021/acs.cgd.3c00388

**Author Contributions**

The experiments were planned by J.d.l.F. and S.R.-G. The Ru films were grown by E.M.T. The magnetite islands were grown by S.R.-G. and J.E.P. The tracks were etched by S.R.-G. and C.F.-G. The experiments were performed by J.d.l.F., S.R.-G., C.G.-M., C.F.-G., L.A., M.A.N., and M.F. J.d.l.F. wrote the manuscript with input and revisions from all the authors.

**Notes**

The authors declare no competing financial interest.

## ■ ACKNOWLEDGMENTS

This work was supported by Grants PID2021-124585NB-C31, RTI2018-095303-B-C51, RTI2018-095303-B-C53, and TED2021-130957B-C54 funded by MCIN/AEI/10.13039/501100011033, by "ERDF A way of making Europe", by the "European Union NextGenerationEU/PRTR", and by the Grant S2018-NMT-4321 funded by the Comunidad de Madrid and by "ERDF A way of making Europe". We thank María Acebrón from IMDEA Nanoscience for her assistance with the optical lithography and etching steps. C.G.-M. acknowledges financial support from grant RYC2021-031181-I funded by MCIN/AEI/10.13039/501100011033 and by the "European Union NextGenerationEU/PRTR".